\documentclass[twocolumn,pre, amsmath, amssymb]{revtex4}

\usepackage{graphicx}
\usepackage{dcolumn}
\usepackage{bm}
\usepackage{braket}
\usepackage{color}

\newcommand{\beq}{\begin{equation}}
\newcommand{\eeq}{\end{equation}}

\newcommand{\bpm}{\begin{pmatrix}}
\newcommand{\epm}{\end{pmatrix}}
\newcommand{\eqname}{Eq.}

\begin{document}


\title{Entropy evaluation sheds light on ecosystem complexity}

\author{Mattia Miotto}
\email{mattia.miotto@roma1.infn.it}
\affiliation{Department of Physics, University ``Sapienza'', Piazzale Aldo Moro 5, 00185, Rome, Italy}
\author{Lorenzo Monacelli}
 \email{lorenzo.monacelli@roma1.infn.it}
 \affiliation{Department of Physics, University ``Sapienza'', Piazzale Aldo Moro 5, 00185, Rome, Italy}%

\date{\today}

\begin{abstract}
Preserving biodiversity and ecosystem stability is a challenge that can be pursued through modern statistical mechanics modeling.
Here we introduce a variational maximum entropy-based algorithm to evaluate the entropy in a minimal ecosystem on a lattice in which two species struggle for survival. 
The method quantitatively reproduces the scale-free law of the prey shoals size, where the simpler mean-field approach fails:
the direct near neighbor correlations are found to be the fundamental ingredient describing the system self-organized behavior. 
Furthermore, entropy allows the measurement of
structural ordering, that is found to be a key ingredient in characterizing two different 
coexistence behaviors, one where predators form localized patches in a sea of preys and another where species display more complex patterns.
The general nature of the introduced method paves the way for its application in many other systems of interest.

\end{abstract}

\maketitle
\section{\label{sec:Introduction}Introduction}

The general formulation of statistical mechanics and information theory opened the way of physics to complex systems. The entropy definition is the basis of both theories. Although the concept of entropy first appeared in thermodynamics, it has been adapted to other fields of study, including  economics, biophysics, and ecology.

In this work, we focus our attention on an ecological system and its entropy, which we show to be pivotal in understanding how the phenotype, the characteristics of an organism resulting from the interaction between its genotype and the environment~\cite{Johannsen_1911,Churchill_1974}, discriminates and provides
information about survival and extinction of species. 
This is a very important effort that must be pursued in order to prevent ecological disasters.

In a very general way, entropy is a property of the distribution function out of which the states of the system have been drawn.
It is the capacity of the data to provide or convey information~\cite{bialek2012biophysics}. Consequently, knowing the entropy allows us to  set limits on the information we can extract from observations and to the predictability of the system.
It has been widely used to study information transport in neural networks~\cite{Strong1998,Nemenman2004} and in  flocks of birds\cite{Bialek2012,GiardinaEntropicEffects,Bialek2001}, complexity and hierarchy in written languages~\cite{Shannon1951} and risk analysis in financial markets~\cite{Kelly1956,Philippatos1972}.
In particular, predictability plays a very important role in economics where the awareness of markets entropy allows us to maximize the investment profits~\cite{Kelly1956}.  
Recently, entropy has been exploited in inference problems, with great results in biological phenomena, such as bacterial growth~\cite{DeMartino2017,2DeMartino2017},
evolution~\cite{Kussell2005} and protein folding~\cite{Weigt2008}.

Here, we show how entropy is crucial also in the context of ecological systems. Ecosystems can be defined as  a community of living organisms in conjunction with the environment~\cite{Tansley1935,Allee1934}, where the latter affects the organisms without being in turn influenced by them~\cite{Chapin2011}.
On the other hand, all living beings within the ecosystem are interdependent, in fact, variations in the size of one population influence all others. This is particularly clear for prey and predator dynamics. 
In fact, if the number of preys in an ecosystem grows, predators will respond to the  supply of available food by increasing their number. The growth of predator number will reduce preys until the system can no longer sustain the predator population. The process has either to attain a steady state or to end in species extinction. 
In order to avoid extinction, both preys and predators need to optimize their phenotypes: predators must, for example, adapt for improving efficiency in hunting  to catch enough food to ensure survival. Prey species, on the other hand, must be proficient in escaping their predators and reproduction; if enough of them are to survive for the species to endure~\cite{cooper2015escaping,Belgrad_2016, Olson_2013}. 
Disturbances, which are perturbations that move the system away from its steady state~\cite{disturbance}, may affect species phenotypes. Such disturbances can originate from changing of environmental variables such as temperature and precipitation or in modifications of the populations, like the appearance/disappearance of a species. 
Besides the theoretical challenge of understanding the behavior of that kind of complex systems, worthy of notice is also the practical importance of predicting the response to perturbations, particularly the ones produced by humans. Relevant cases are the fight against parasites in agriculture~\cite{Kindlmann2015} and the perturbations in sea populations due to fishing activities~\cite{Bax1998}.
From the groundbreaking works of Lotka~\cite{Lotka} and Volterra~\cite{volterra} ecosystem modeling has been addressed in various ways, from sets of differential equations~\cite{Holmes_1994, Kuto_2004,Ferreira_2013} to simulations on lattice~\cite{Satulovsky_1994,Lipowski_1999,Dobramysl2018}.    

Taking inspiration from the work of Dewdney~\cite{dewdney} we modeled the simplest nontrivial ecosystem in which two species struggle for survival. 
Sharks (predators) and fishes (preys) occupy the nodes of the toroidal 2D lattice; they can move, reproduce and hunt. The rules of the model resemble the ones  described by Mobilia et al.~\cite{Mobilia2006} and are introduced in Sec.~\ref{sec:mod_def}.
This system has been extensively studied and several critical behaviors have been observed~\cite{Antal2001,Tuber2011}.
 
Measuring the entropy of this kind of complex system without any explicit expression for a \emph{prior} probability distribution is very challenging~\cite{nemenman2002entropy}. In fact, the Shannon definition of entropy relies on the system probability distribution $P$, which depends on all the degrees of freedom of the system~\cite{shannon_game}.
The entropy measurement requires a fine sampling of this function, becoming not affordable even for small lattices.

The Maximum Entropy (MaxEnt) technique has been developed in order to obtain an approximation for the probability distribution.
 Given a set of observables ${x}_i$ that partially describes the inquired system, MaxEnt algorithm allows finding the less structured probability distribution that reproduces the chosen set of the real system observables.
This technique was firstly introduced by Jaynes~\cite{Jaynes1957} in 1957, but it reached an outstanding interest only recently thanks to the availability of a huge amount of experimental and numerical data.
MaxEnt has been successfully applied to countless problems, both in equilibrium and out-of-equilibrium systems~\cite{Cavagna2014}; among them, the prediction of protein amino acid contacts~\cite{Weigt2008,Mora2010} and the analysis of neural networks~\cite{Schneidman2006} are of particular interest. 
However, even if MaxEnt enables one to extract an analytic expression for the probability distribution, computing the associated entropy is still a major issue~\cite{GiardinaEntropicEffects}.

In sec.~\ref{sec:entropy:eval} we outline how it is possible to obtain the exact entropy of the MaxEnt probability distribution taking advantage of all the data generated during the convergence of the algorithm, without any further time-consuming computation. Furthermore, in sec. \ref{sec:least:maximum:entropy}, we introduce a \emph{least} entropy principle that justifies the use of the MaxEnt distribution as a truncation of a series that converges toward the real entropy of the system.
Availing the variational principle, the approximation on the resulting entropy is of second order.
While the method is formally derived for dealing with equilibrium systems, in sec.~\ref{sec:eco:anal} we apply it to the study of an ecosystem. 
In sec.~\ref{sec:Discussion} we examine the insights the entropy provides and discuss its limitations in the case of a non-equilibrium steady-state system, like the one we treat.  
Although we apply this method only to the introduced ecosystem, it is very general: it can be used whenever it is possible to define a probability distribution on a site model.

\subsection{Maximum Entropy}\label{sec:ME} 
The MaxEnt framework we are going to discuss can deal with any stochastic process defined on a graph composed of $M$ nodes. Each node is in one out of $q$ possible states.
If we indicate with $P$ the probability that the system is in a given configuration the entropy is defined accordingly to Boltzmann-Shannon as:
\beq
S[P] = -\braket{\ln P}_P,
\eeq
where $\braket \cdot_P$ indicates the average over the $P$ probability distribution.

The standard MaxEnt algorithm consists in maximizing $S[P]$ with respect to $P$ in presence of a set of $N$ constraints.
The restraints are the set of observables $\left\{x\right\}_{i=1}^N$ that best describes the system; in other words, some degrees of freedom are fixed and the entropy is maximized
among the remaining ones in order to have the broadest possible probability function. 
Each observable $x_i$ is a generic function that associates any possible configuration of the system to a real number. Note that the index $i$ identifies the observable in the set of constraints, not the specific site on the graph, as the $x_i$ observable can be a function of more than one node, e.g. the average dimension of clusters of nodes in the same state.
 
Defining the auxiliary Lagrange functional $\Phi$ as:
\beq
\label{eq:lagmul}
\Phi[P, \lambda_1, \cdots, \lambda_n] = -S[P] + \sum_{i = 1}^N \lambda_i \braket{x_i}_P,
\eeq
the correct constrained maximum entropy distribution $P^*$ is found solving the set of equations:
\begin{subequations}
\label{eq:max:ent:before}
\begin{eqnarray}
\label{eq:lagrange:min}
\frac{\delta \Phi}{\delta P^*} &=& 0\\
\braket{x_i}_{P^*} &=& \bar{x_i} \;\;\;\; \forall i = 1, \dotso, N
\end{eqnarray}
\end{subequations}
where $\bar x_i$ is the measured expected value of $x_i$.
\eqname~\eqref{eq:max:ent:before} is very hard to solve, even numerically, since $P$ depends still on $q^{M}$ variables.
$P$ can be used to define an auxiliary effective Hamiltonian, according to the Boltzmann definition:
$$
P(\vec \sigma) = \frac{e^{- H(\vec \sigma)}}{Z},
$$
where $\vec \sigma$ represents the configuration and $Z$  the partition function that normalizes $P$.

\eqname~\eqref{eq:lagrange:min} is solved by the Hamiltonian~\cite{Jaynes1957}:
\beq
\label{h:sol}
H =  \sum_{i = 1}^N \lambda_i x_i,
\eeq
where $\lambda_i$ are fixed so that the expectation values of $x_i$ respect the bounds over the real observables.

The values of $\lambda_i$ can be obtained analytically only in very few cases. The simplest one is the mean-field solution, where only one-body observables are constrained, e.g. the numbers of nodes in each state. In the latter case the number of observables is equal to the number of states $q$ and the Lagrange multipliers that satisfy the imposed constraints are:
\beq
\label{eq:mf_sol}
\lambda_i = -  \ln \left(\frac{\bar {x_i}}{M}\right).
\eeq
The passages to prove \eqname~\eqref{eq:mf_sol} are sketched in appendix~\ref{app:MFME}.
In this case, the entropy per node is:
\begin{subequations}
\begin{eqnarray}
\label{eq::fano}
S' &=& - \sum_{i = 1}^q \left(\frac{\bar {x_i}}{M}\right)\ln \left(\frac{\bar{x_i}}{M}\right),\\
S &=& M  S'.
\end{eqnarray}
\end{subequations}
This entropy evaluation corresponds to the standard one obtained by the Shannon-Fano algorithm~\cite{Shannon1948,Fano1949}. This
is a `Hartree-Fock' theory of the complex system, where the entropy is maximized using topology-independent Hamiltonians only.
This framework paves the way to a more precise entropy computation.

\section{Entropy algorithm}\label{sec:alg}
The general solution of \eqname~\eqref{eq:max:ent:before} has been matter of discussion~\cite{Lockless_1999,Socolich_2005,Russ_2005,Schneidman2006,2007arXiv}.
On the way of Bialek and Ranganathan~\cite{2007arXiv}, we introduce an auxiliary function ${\tilde\chi}^2$ whose global minimum coincides with the solution.

\beq
\label{chi2}
{\tilde\chi}^2 = \sum_{i = 1}^N W_i \left(\braket{x_i}_H - \bar x_i\right)^2,
\eeq 
where $\braket{x_i}_H$ is the average of the $x_i$ observable computed using the trial Hamiltonian $H$ while $\bar x_i$ is the average over the real observable evaluated by the data. $W_i$ are coefficients that do not affect the minimum of the $\chi^2$, however, if wisely chosen, may accelerate the convergence process.
The ${\tilde\chi}^2$ function defined in \eqname~\eqref{chi2}, in the minimum, is a random distributed Pearson variable only if all the observables are  independent ofeach other and the $W_i$ correspond to the inverses of the variances. This is not true in most MaxEnt applications, e.g. in the mean field case where  the sum of the $x_i$ is fixed to $M$.

We propose the introduction of a corrected $\chi^2$ that  takes into account linear correlations.
\beq
\chi^2  = \sum_{\substack{i = 1\\ j = 1}}^N \left(\bar x_i - \braket{x_i}_H\right) \left(\Sigma^{-1}\right)_{ij} \left(\bar x_j - \braket{x_j}_H\right),
\label{chi2:new}
\eeq
where $\Sigma$ is the covariance matrix between the $N$ chosen observables. 
It must be stressed that \eqname~\eqref{chi2:new} is still not a true $\chi^2$ variable since we corrected only linear correlations. Moreover, eigenvectors of $\Sigma$ will be uncorrelated, but not necessarily independent. However, the linear approximation for correlations between observables has a long history in statistical analysis~\cite{Pearson_1901,Hotelling_1933} and usually leads to very good results.

\eqname~\eqref{chi2:new} is well defined only if $\Sigma$ can be inverted.
Diagonalizing the covariance matrix $\Sigma$, we can restrict the minimization only in the subspace spanned by its eigenvectors whose eigenvalues are greater than zero. Using these eigenvectors $y_i$ as a basis, the Pearson $\chi^2$ can
be redefined as:
\begin{subequations}
  \begin{eqnarray}
    \chi^2 &=& \sum_{i = 1}^{N'} \frac{\left(\bar y_i - \braket{y_i}_H\right)^2}{{\tilde\sigma_i}^2},
\label{eqn:chi2:final}\\
N' &=& N - \dim{\ker{\Sigma}}, \\
y_i &=& \sum_{j=1}^{N} S_{ij} x_j,
\label{eigenval}
  \end{eqnarray}
\end{subequations}

where $\tilde\sigma_i^2$ is the $i$-th eigenvalue of the $\Sigma$ matrix and $S$ is a $N'\times N$ matrix that diagonalizes $\Sigma$.
The $\dim{\ker{\Sigma}}$ indicates the dimension of the $\Sigma$ kernel.

The gradient of \eqname~\eqref{chi2:new} can be computed as follows:

\begin{subequations}
  \begin{eqnarray}
\frac{\partial \chi^2}{\partial \lambda_k} &=& -2\sum_{i = 1}^{N'} \frac{(\bar y_i - \braket{y_i}_H)}{\tilde\sigma_i^2} \frac{\partial \braket{y_i}_H}{\partial \lambda_k},\\
\frac{\partial \braket{y_i}_H}{\partial\lambda_k} &=& \sum_{j = 1}^{N} S_{ij} \frac{\partial \braket{x_j}_H}{\partial\lambda_k},\\
\frac{\partial {\braket{x_j}_H}}{\partial \lambda_k} &=& -\sigma^{MC}_{jk}.
  \end{eqnarray}
\end{subequations}
where  $\sigma^{MC}_{jk}$ is the covariance matrix between observables $x_j$ and $x_k$ for the current Hamiltonian:
\beq
\sigma_{jk}^{MC} = \braket{x_jx_k}_H - \braket{x_j}_H\braket{x_k}_H.
\eeq
The final expression of the gradient is:
\beq
\frac{\partial\chi^2}{\partial\lambda_k} = 2  \sum_{i = 1}^{N'} \frac{(\bar y_i - \braket{y_i}_H)}{\tilde\sigma_i^2}\sum_{j = 1}^N S_{ij}\sigma^{MC}_{jk},
\label{eqn:gradient:new}
\eeq
or, equivalently, in a compact form:
\beq
\vec \nabla \chi^2 =  2S^\dagger  \sigma_{MC}' \left(\frac{\vec {\Delta y}}{\sigma^2}\right),
\eeq
where $\sigma_{MC}'$ is the Monte Carlo covariance matrix in the non singular subspace and $\vec{\Delta y}/\sigma^2$ is the  vector:
\beq
\left(\frac{\vec {\Delta y}}{\sigma^2}\right)_i = \frac{\bar y_i - \braket{y_i}_H}{\tilde\sigma_i^2}.
\eeq

The minimization of \eqname~\eqref{chi2:new} can be initialized by the mean-field solution~\eqref{eq:mf_sol}, choosing zero for each $\lambda_i$ associated with a non topology-independent observable. \eqname~\eqref{eqn:gradient:new} ensures that any standard gradient-based minimization algorithm can be used.

Moreover, in order to fasten the convergence\cite{Nash1985}, it is possible to  derive the expression of the Hessian matrix in the minimum and utilize it  as a precondition on the minimization:
\beq
D_{hk} = \left.\frac{\partial^2 \chi^2}{\partial\lambda_k\partial\lambda_h} \right|_{\vec \nabla \chi^2 = 0},
\eeq
\beq
D_{hk} = 2 \sum_{i = 1}^{N'} \frac{1}{\tilde{\sigma^2_i}} \sum_{\substack{j = 1\\ l=1}}^N S_{ij}S_{il}\sigma^{MC}_{jk} \sigma^{MC}_{hl},
\eeq
or, equivalently:
\beq
D = 2 \sigma_{MC} S^\dagger \Sigma^{-1} S \sigma_{MC}.
\eeq

\subsection{Entropy evaluation}\label{sec:entropy:eval}
In usual MaxEnt implementations, minimization data are wasted and information about the system is inferred solely from the final probability distribution. 
Here we show how to recycle the whole minimization procedure to infer the entropy of the system.
In fact, computing the entropy directly from the converged probability distribution is a very challenging task. However, entropy can be obtained from  an adiabatic integration through the minimization path of the Hamiltonians.
The values of the observables during the minimization can be used to obtain a measurement of the entropy of the system
without  any further Monte Carlo computation.

To compute entropy, it is convenient to define, as done for the effective Hamiltonian, an auxiliary function that is equivalent to the Helmholtz free energy:
$$
F = -\ln Z,
$$
which can be computed through a thermodynamic integration along the minimization path. Then, 
Entropy  is obtained by  inverse Legendre transformation from the auxiliary free energy of the system.
Even if the free energy is not well defined (the energy is defined up to a constant), the entropy is.

The free energy at the final value of the minimization is:
\beq
F(\xi = 1) = F_0 + \int_0^1 \frac{dF}{d\xi} d\xi,
\eeq
where $\xi$ is a variable that parametrizes the path of the Hamiltonian during the minimization. The $F_0$ value is the
free energy at the starting condition, that is the non-interacting system.
\beq
F_0 = -M \ln \left(\sum_{j = 1}^q e^{-\beta \lambda_j(0)}\right),
\eeq
\beq
F_\xi = - \ln Z_\xi,
\eeq
\beq
\frac{dF}{d\xi} = \braket{\frac{d H_\xi}{d\xi}}_\xi.
\eeq

The integral can be done parametrizing the Hamiltonian as:
\beq
H_\xi = \sum_{i = 1}^N \lambda_i(\xi) x_i ,
\eeq
where $x_i$ are the observables while  $\lambda_i$ are both the Lagrangian multipliers of the MaxEnt algorithm and the parameters through which the $\chi^2$ function is minimized.
Therefore we get:
\beq
\frac{dF}{d\xi} = \sum_{i = 1}^N \frac{d\lambda_i}{d\xi} \braket{x_i}_\xi,
\eeq
so that  only the averages of the observables during the minimization are required in order to compute the free energy:

\begin{subequations}
  \begin{eqnarray}
    F &=& \braket H - TS,\\ S &=& \frac{\braket H - F}{T}.
  \end{eqnarray}
\end{subequations}
Fixing  $T = 1$, we obtain:
\begin{widetext}
\beq
\label{eq::entropy_cal}
S[\lambda_i(\xi)] = M \ln\left(\sum_{j = 1}^q e^{-\lambda_j(0)}\right) + \sum_{i = 1}^{N}\left[ \lambda_i(1)\braket{x_i}_1 - \int_0^1 d\xi \frac{d\lambda_i}{d\xi}(\xi) \braket{x_i}_\xi\right].
\eeq
\end{widetext}

The only required quantities are the Hamiltonian during the minimization, i.e. the $\lambda_i(\xi)$, and the
values of the observables $\braket{x_i}_\xi$,  both  already computed during the minimization. In addition, if all the configurations generated during the $\chi^2$ minimization are stored, the importance sampling (IS) can be used to interpolate between different Monte-Carlo points, providing a very good sampling of the integral. IS implementation for the minimization is discussed in Appendix~\ref{app:IS}. 

\subsection{Least maximum entropy principle}
\label{sec:least:maximum:entropy}
Entropy can be defined in the framework of a least principle.
The MaxEnt approach finds the probability distribution that maximizes the entropy on the subset  where the expected values of the observables $x_i$ are constrained. The entropy $S_{ME}$ associated with the MaxEnt probability distribution is greater than the true entropy $S_{real}$ of the system since the true  $P$ lies in  the chosen subset.
\beq
S_{real} \le S_{ME}.
\label{eq:least:maximum}
\eeq

Moreover, $S_{ME}$ decreases whereas new constraints are added due to a contraction of the probability distribution space.
In Appendix~\ref{app:least} we provide a rigorous proof of the fact that a set of constraints that ensures \eqname~\eqref{eq:least:maximum} to become an equality exists.
The true entropy of the system is then the least maximum entropy of all possible choices of the constraints.

Just like any variational least energy principle in physics, from Hartree-Fock to \emph{density functional theory} (DFT), the energy (entropy in our case) and its derivatives are the targets of the theory, while the wave functions (probability distributions)
are \emph{side-effects}. We want to remark that the error on the entropy due to the limited number of constraints is of second order, while the resulting probability distribution is affected by a first order error.

\section{Ecosystem analysis}\label{sec:eco:anal}
Although the so far introduced method  is quite general, we discuss its implication in ecosystems.
In particular, we analyze a two dimensional model on a regular 2D lattice of edge $L$ (number of nodes $M = L^2$), whose sites can either be empty or  occupied by a fish or a shark. In the application of the MaxEnt algorithm, we limited the constraints $x_i$ introduced in \eqname~\eqref{eq:lagmul} to the numbers of preys and predators and  near-neighbor fish-fish, shark-shark and fish-shark couples.
The corresponding MaxEnt Hamiltonian describes a three-state Potts model~\cite{Potts_1952}.

\subsection{Model definition}\label{sec:mod_def}

Along the lines of Dewdney~\cite{dewdney} and Mobilia et al.~\cite{Mobilia2006}, we modeled a minimal ecosystem composed of two species interacting each other as a 2D lattice model. 
Each site can be occupied either by the environment or a fish or a shark, respectively represented with the integers 0,1,2.
At every time step, fishes can move, breed or remain still with probability $p_f^m$, $p_f^f$ and $1 - p_f^m - p_f^f$. Sharks can move ($p_s^m$) or remain still ($1 - p_s^m$). Furthermore, sharks eat fishes whenever they step into a cell occupied by a prey. In this case, sharks can reproduce with probability $p_s^f$. If a shark does not eat a fish during its round, it can die with probability $p_s^d$.    
This set of rules defines a Markovian process described by the transition matrix $\Pi(\vec \sigma_i \rightarrow \vec\sigma_j)$. It gives the probability of the system to transit from 
the $\vec \sigma_i$ to the $\vec \sigma_j$ state, where $\vec \sigma_x$ identifies the values of all sites in the lattice.
The probability to find the system in the $\vec \sigma_i$ state at the $t + 1$ time step is defined by:
\beq
P(\vec\sigma_i, t+ 1) = \sum_{\vec \sigma_j} \Pi(\vec\sigma_j \rightarrow \vec\sigma_i) P(\vec\sigma_j, t).
\label{eq:trans:mat}
\eeq
Studying the time evolution of $P(\vec \sigma, t)$ through Dynamical Monte Carlo simulations (see appendix~\ref{app:dyn:mc}) it is 
possible to
assess that the system reaches a steady state distribution:
\beq
P(\vec\sigma_i) = \sum_{\vec\sigma_j} \Pi(\vec\sigma_j \rightarrow \vec\sigma_i) P(\vec\sigma_j).
\label{eq:equilibrium}
\eeq

Depending on the choice of the parameters, the system that we named EcoLat (\textit{Eco}system on \textit{Lat}tice) presents three different states: i) fish saturation due to the extinction of sharks; ii) life extinction, where sharks eat all fishes and then extinguish. iii) Non-Equilibrium Steady-State (NESS), in which fish and shark  densities fluctuate around a constant value. 

The probability distribution can be defined by extracting configurations from the evolving system after a transient time. Sampling configurations sufficiently distant in time it is possible to introduce an ergodic hypothesis (appendix~\ref{app:ergodic}), i.e. the so defined probability distribution is equivalent to the one of an infinite ensemble of independent systems.
In this framework, the entropy becomes a well-defined quantity. 

It is important to notice that such formulation neglects the time-correlations of the evolving configuration, i.e. it disregards the flux of probability  distribution that uniquely characterizes the generic non-equilibrium steady-state~\cite{Zia2007}.

\subsection{MaxEnt distribution benchmark}
\begin{figure*}
  \includegraphics[width=\textwidth]{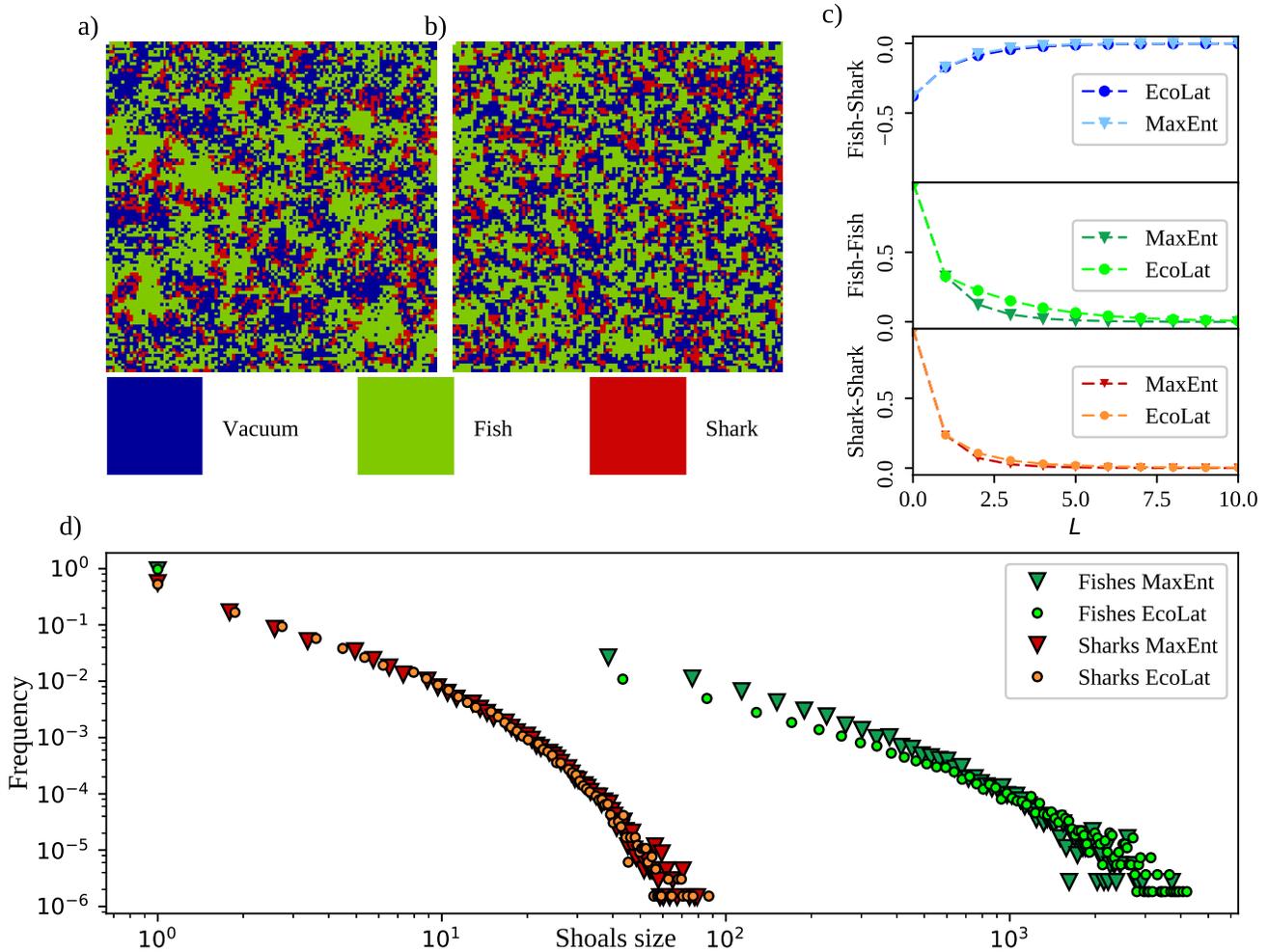}
  \caption{(Color online) Comparison between the modeled ecosystem (EcoLat)  and the Maximum Entropy (MaxEnt) result.
  \textbf{a} Representation of an EcoLat snapshot in the steady-state regime. Fishes are colored in green, sharks in red, while blue represents the environment. Phenotypic parameters are reported in Appendix~\ref{app:par}. \textbf{b} A configuration extracted from the  MaxEnt probability distribution obtained by constraining the numbers of fishes, sharks and near neighbor couples.  
  Both simulations ran on a lattice of  edge $L=110$. 
    \textbf{c}. Spatial correlation functions of fish-shark (top), fish-fish (medium) and shark-shark (bottom).
    \textbf{d}. Shoal size distribution of fishes (green shades) and sharks (red shades)  computed on EcoLat  (circles) and MaxEnt  (triangles) configurations. Fishes are power-law distributed, while sharks exhibit   an exponential decay.\label{fig::SF}}
\end{figure*}

The configurations extracted from EcoLat are used to evaluate the constraints of the MaxEnt distribution (the number of fishes, sharks and near neighbor couples).
We found a good agreement between EcoLat and MaxEnt distributions (\figurename~\ref{fig::SF}).

In \figurename~\ref{fig::SF}(\emph{a-b}) two sample configurations are compared. The general aspect of the system is well reproduced by near neighbors MaxEnt, except the shape of the shoals that exhibits some differences. This is reflected by the spatial correlation functions in~\figurename~\ref{fig::SF}(\emph{c}), computed as the Pearson coefficient:
\beq
f_{ij}(x) = \frac{\braket{\sigma_i(X)\sigma_{j}(X+x)} - \braket{\sigma_i(X)}\braket{\sigma_{j}(X +x)}}{\sqrt{(\braket{\sigma_i(X)^2} - \braket{\sigma_i(X)}^2)
(\braket{\sigma_j(X)^2} - \braket{\sigma_j(X)}^2)}}, 
\label{eq:corr:coeff}
\eeq
where $\sigma_i(X)$ is one if the site $X$ is occupied by the $i$-th species; note that $f_{ij}$ does not depend on $X$ thanks to the translational symmetry of the system. 
MaxEnt approximation, although it maintains the qualitative agreement, predicts lower fish-fish spatial correlation at larger distances. This does not affect the cluster size distribution, see \figurename~\ref{fig::SF}(\emph{d}), that decays  with the same slope both in EcoLat and MaxEnt.
The reason can be understood by looking at the snapshot of the configuration in \figurename~\ref{fig::SF}(\emph{a-b}), where fishes create shoals of similar size but having  more roundish shapes in EcoLat than in MaxEnt configuration, explaining the  higher spatial correlations even if shoals exhibit the same size distributions.
This is a general feature of the system independent on the choice of the phenotypes. It unveils that, in the dynamical steady-state, 
fishes cooperatively interact beyond near-neighbors, while all other interactions seem operating on near-neighbor sites.

The power-law decay in EcoLat shoal size distribution (\figurename~\ref{fig::SF}\emph{d}) has been already observed~\cite{Sutherland1994} and assigned to a self-organized behavior of the system, moreover, it seems to be a general characteristic of several spatial ecology models~\cite{Pascual2002}.

The MaxEnt distribution is very close to a critical point. This can be checked by introducing a parameter $T$ in analogy to the Boltzmann temperature, however, the $T$ dependence alone is not an evidence of the criticality in the original system~\cite{Mastromatteo2011}.



\subsection{Entropic curves}
Thanks to the method introduced in \eqname~\eqref{eq::entropy_cal}, it is possible to compute the entropy of the model and to shed light on several features of the ecosystem.

\begin{figure*}
\centering
\includegraphics[width=\textwidth]{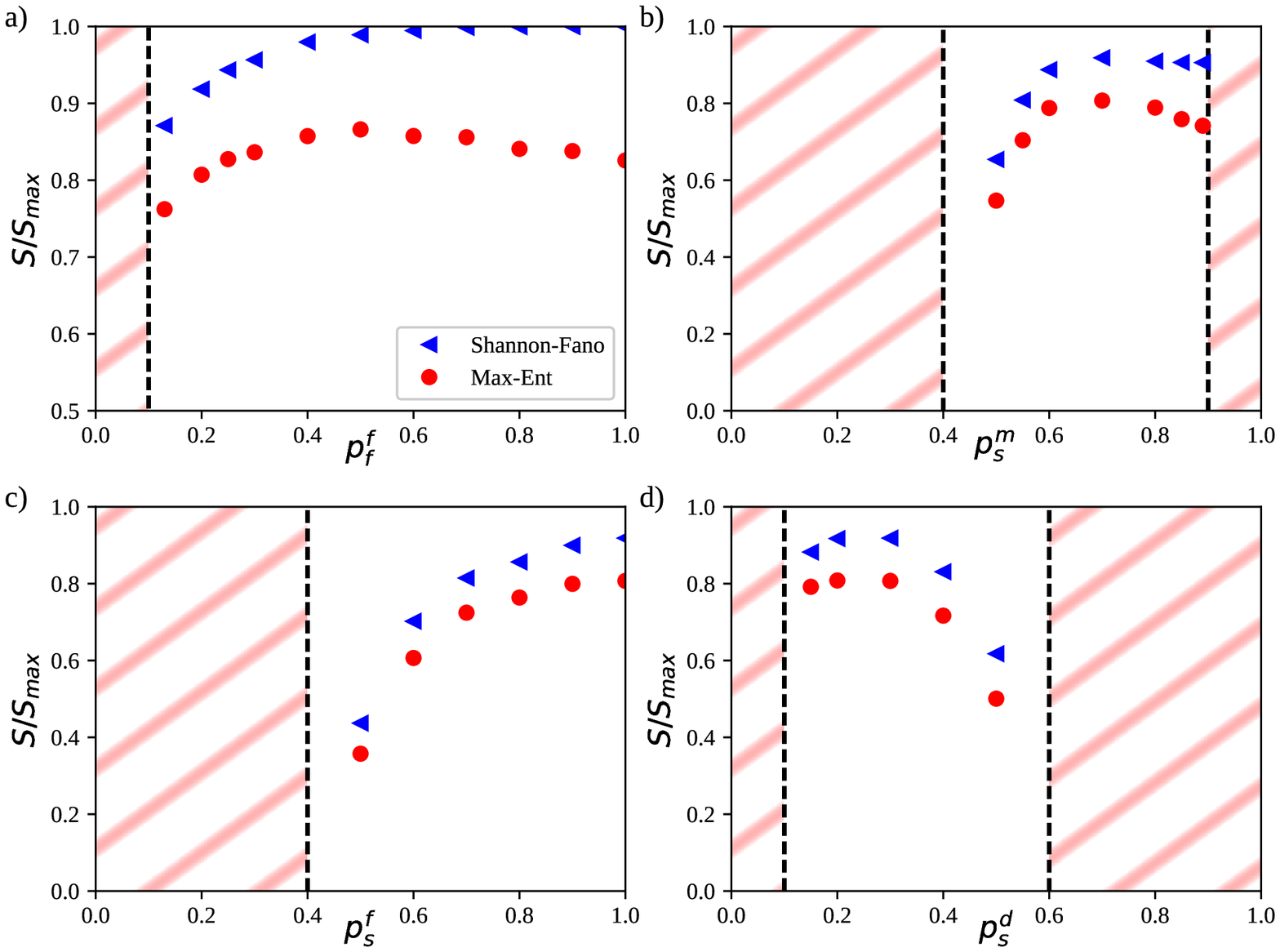}
\caption{(Color online) Entropy per site as a function of species phenotypes. Blue triangles indicate the Shannon-Fano entropy~\eqref{eq::fano} while red circles the MaxEnt entropy~\eqref{eq::entropy_cal}. The ranges of parameters  that lead species to extinction are underlined by obliques lines. \textbf{a)} Entropy vs fish breading probability. A qualitative difference in the entropy behaviors appear when $p_f^f > 0.5$; MaxEnt entropy starts decreasing while Shannon-Fano one saturates to the maximum value.
\textbf{b)} Entropy vs shark mobility. Also in this case a similar difference in behaviors manifests in the region $0.7< p_s^m < 0.9$. These differences  outline that structural ordering is occurring in the system. 
  \textbf{c)} Entropy vs shark filiation.  \textbf{d)} Entropy vs shark mortality.
In these two cases, the Shannon-Fano approximation grasps qualitatively well the entropy trends, although the numerical values are overestimated.
\label{fig:entropy}}
\end{figure*}

In \figurename~\ref{fig:entropy} we report, as a function of the species relevant phenotypes, 
the entropy per site  of the system normalized by its maximal value $\ln{3}$. The entropy has been measured both through the mean-field Shannon-Fano algorithm~\eqref{eq::fano} and with the new approach based on the MaxEnt calculation of \eqname~\eqref{eq::entropy_cal}. The MaxEnt entropy estimation is always lower than the mean-field result, as expected due to the variational nature of the least-entropy-principle.
Regions of phenotype values that lead the system to extinction are filled with obliques lines.
\figurename~\ref{fig:entropy}(\emph{a-b}), showing entropy curves as a function of $p_f^f$ and $p_s^m$ respectively, manifest  qualitative differences between Shannon-Fano and MaxEnt entropy trends. In particular, while Shannon-Fano predicted entropies reach a plateau whenever $p_f^f \gtrsim 0.5$ or $p_s^m \gtrsim 0.7$, MaxEnt ones display a maximum around those values. 
The increasing difference between Shannon-Fano and MaxEnt entropy is a clear sign that  structural ordering is occurring since MaxEnt entropy is the exact entropy of the Potts-like Hamiltonian that takes into account spatial correlations even beyond near-neighbor ones (\figurename~\ref{fig::SF}\emph{c}).  	 

This feature, particularly visible in \figurename~\ref{fig:entropy}(\emph{b}), is related to the formation of waves of predator and preys (see snapshots in \figurename~\ref{fig:PTD}\emph{b}). 

Furthermore, extrapolation of \figurename~\ref{fig:entropy}(\emph{b-c-d}) entropy curves manifests a sharp point in $p_s^m = 0.4$, $p_s^f = 0.4$ and $p_s^d = 0.6$, which are a peculiar sign of a second order phase transition between coexistence and extinction. On the contrary, when $p_s^m = 0.9$ or $p_s^d = 0.1$ entropy displays a sudden jump into the extinction phase. This transition is due to a finite size effect~\cite{PhysRevE.80.021129,PhysRevE.85.051903}: increasing the lattice size the probability of the system to extinguish in a fixed time becomes sharper as a function of phenotypes but the transition threshold drifts to the bond value $p_s^m = 0$ in the thermodynamic limit.

\subsection{Discussion}\label{sec:Discussion}
The measurement of disorder provides a new insight into the ecosystem. It allows us to recognize the second order phase transition near predator extinction threshold,
to characterize the self-organized behavior of prey shoals, and 
to unveil the increase of structural ordering the system acquires improving the predator hunting capability. 
Here we discuss these findings following the ecosystem behavior while tuning the shark mobility ($p_s^m$) as in \figurename~\ref{fig:entropy}(\emph{b}) since it is a particularly explicative parameter of the model.
Increasing the shark mobility the system passes from an absorbing state where the lattice is crowded with preys to a phase in which sharks start appearing in small shoals swimming in a sea of fishes. This is a critical point known in literature~\cite{Antal2001,Tuber2011,Dobramysl2018} and it was characterized by its dynamical properties, where predator population decays in time with a power-law and it belongs to the directed percolation universality class. The criticality of this point 
is reflected in the entropy behavior which manifests a sharp point. 

This phase transition, marking the passage between species coexistence and extinction, happens where the entropy reaches zero (sharp points in  \figurename~\ref{fig:entropy}(\emph{a-b-c}) at $p_s^m = 0.4$, $p_s^f = 0.4$ and $p_s^d = 0.6$), and can be attained diminishing
the shark hunting ability as well as their fertility or increasing their mortality.
The zero entropy of this critical point can be explained in a Shannon-Fano framework, in fact, the predominance of preys unbalances the average numbers of sharks and fishes.

\begin{figure*}
\centering
\includegraphics[width=\textwidth]{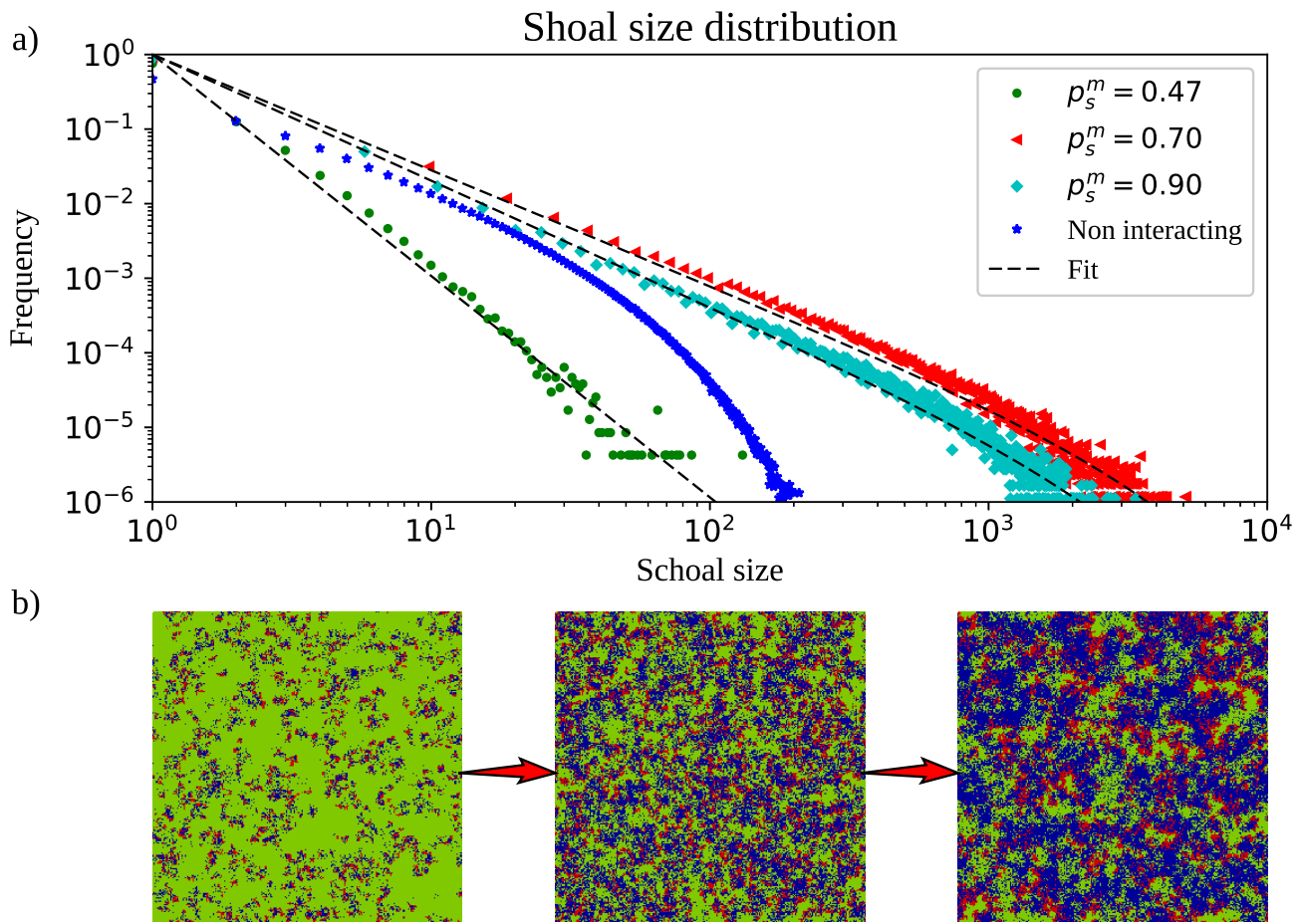}
\caption{(Color online) \textbf{a} Distributions  of prey shoal sizes for 
three different values of predator mobility (all parameters are reported in Appendix \ref{app:par}).
Distributions for $p_s^m= 0.47$,$0.70$,$0.90$ are shown in green dots, red triangles and cyan diamonds respectively. Blue stars represent the distribution of the independent-site model (Noninteracting) with the fish density fixed to match the EcoLat one with $p_s^m = 0.7$.
The EcoLat distributions have been fitted with a function $x^{-\gamma}\exp(x / \xi)$. The lower mobility case  has $\xi =  \infty$ and $\gamma = 2.97 \pm 0.04$, medium mobility has
$\xi =  3450\pm 220$ and $\gamma = 1.55 \pm 0.01$, while high shark motility gives a fish shoal distribution with $\xi = 2440\pm 180$ and $\gamma = 1.69\pm 0.01$. \textbf{b} Snapshots taken from the EcoLat steady state distribution while tuning the shark mobility $p_s^m$ from 0.47 (left) to 0.90 (right). Fishes are depicted in green, sharks in red, empty sites in blue.
\label{fig:PTD}}
\end{figure*}

Moving away from the aforesaid criticality, fish clusters  acquire a power-law distribution~\cite{Pascual2002} well described by the MaxEnt approximation (\figurename~\ref{fig::SF}).
Very interestingly, this kind of distribution cannot be simply explained in a Shannon-Fano framework. 
In fact, if we simulate an independent size model, fixing the densities as the real EcoLat ones, we obtain  
 the cluster size distribution of \figurename~\ref{fig:PTD}(\textit{a}) that does not match with the EcoLat one. Moreover, the Fisher exponent of the EcoLat distribution varies in the studied parameter ranges between 1.5 and 3 and does not match with the fixed 2.05 one of the independent-site model at the percolation threshold (ordinary percolation). It is worth to notice that the observed power-law is completely accounted for and described by near-neighbor interactions, as the MaxEnt approach is able to quantitatively reproduce it (\figurename~\ref{fig::SF}\emph{d}).
Furthermore, the MaxEnt Hamiltonian is very close to a critical point. Even if this feature alone is not a signature that the real EcoLat is itself close to
a criticality~\cite{Mastromatteo2011}, the MaxEnt power-law cluster distribution is related to this criticality. 
Is the power-law distribution a marker of self-organized criticality in the EcoLat model? The excellent accordance with the critical MaxEnt power-law seems to
support this hypothesis. 


Increasing $p_s^m$ from the directed percolation critical point we see the disorder growing (first two snapshots of \figurename~\ref{fig:PTD}\emph{b}). At first, the system starts getting rid of fish dominance (i.e. it moves away from direct percolation critical point) simply increasing the number of predators. Both the higher number of sharks and their increased motility make the ecosystem drift toward more disordered configurations (both Shannon-Fano and MaxEnt entropies increase). At a certain point, the interactions between species dominate and the system passes to a regime where it starts regaining order.
From \figurename~\ref{fig:entropy}(\textit{b}) and \figurename~\ref{fig:PTD}, we see that entropy discriminates these two distinct dynamical behaviours of species at coexistence: the one with a predominance of fishes and sharks grouping in small shoals, and the other where both  preys and predators form elaborate shoals, characterized by a spreading wave-like fronts of fishes and sharks~\cite{Mobilia2007p,Dobramysl2018} with predators surrounding the shoals of preys (\figurename~\ref{fig:PTD}\emph{b}).
 Very interestingly, this crossover is clearly characterized by a decreasing MaxEnt entropy with respect to a constant Shannon-Fano one (\figurename~\ref{fig:entropy}\emph{b}), remarking the structuring of the system and the impossibility of grasping this behavior just considering the mean-field approximation. The progressive re-achievement of order can be visualized by looking at the last two EcoLat snapshots in the steady-state regime (\figurename~\ref{fig:PTD}\emph{b}). Notably, this new structural order is
very different from the mean-field order close to the shark extinction threshold (last vs first snapshots of \figurename~\ref{fig:PTD}\emph{b}). 
The structural order can be measured as the difference between Shannon-Fano and MaxEnt entropy. In fact, Shannon-Fano order is a mean-field quantity that
does not depend on the disposition of species in the lattice, while MaxEnt entropy considers the order resulting from all possible correlations reproduced by a three-state Potts model. An analogy with the Ising model (where the MaxEnt algorithm is exact) gives a clearer picture: in the overcritical region, $T > T_c$ and no external magnetic field, the Shannon-Fano entropy is always at its maximum value since there is an equal number of spin up and spin down. However, the true entropy decreases as $T \rightarrow T_c$ since ordered spin domains appear.

Finally, another transition between coexistence and extinction is reached by further increasing the shark mobility (as well as decreasing the fish fertility and shark mortality). Contrariwise to the first critical point, the entropy does not continuously go to zero in correspondence of the phase transition, but abruptly jumps to zero  (\figurename~\ref{fig:entropy}).
This is a finite-size effect~\cite{PhysRevE.80.021129,PhysRevE.85.051903} that disappears when $L\to\infty$. It is worth to notice that, at fixed $L$, it is possible to continuously tune the phenotype in order to get a steady-state metastable phase even in the extinction region that lives until a sufficiently large stochastic fluctuation causes a brutal extinction. This is very worrying; in fact, it is difficult to predict since no precursors are present and can bring to an ecological catastrophe.

\section{Conclusions}
Entropy measurements in complex systems have always been challenging. MaxEnt is a powerful tool to obtain an estimation of the probability distribution of the system from simulations or experimental data. Until now, the information provided by the intermediate steps of the MaxEnt solution was wasted. 
We introduce a way to recycle it in order to directly evaluate the entropy of the system without any further time-consuming computation~(\eqname~\ref{eq::entropy_cal}). Thanks to the MaxEnt nature of the probability distribution, a variational principle for entropy evaluation of the real system  can be formulated, which ensures that the obtained entropy is always greater or equal to its true value. Moreover, \eqname~\eqref{eq::entropy_cal} is quite general, allowing to compute entropy where it is possible to formulate a MaxEnt algorithm.

Among many possible applications, the knowledge of entropy in ecological systems plays a pivotal role describing the complexity due to the phenotype variability. In the studied prey-predator ecosystem, it sheds new light on the self-organizing behavior of preys. 
The direct near-neighbor correlations used in our MaxEnt approach are found to be the fundamental ingredient in this self-organized behavior: The MaxEnt Hamiltonian quantitatively reproduces the scale-free behavior of the prey shoals size, where the simpler mean-field approach fails. Furthermore, entropy allows the measurement of
structural ordering, that is found to be a key ingredient in characterizing the crossover between two different 
coexistence behaviors, one where predators form localized patches (dominated by mean-field disorder) and another where predators chase preys in spreading prey-predator fronts (dominated by structural order).

As a matter of fact, the entropy curves reported in \figurename~\ref{fig:entropy} are a powerful tool to investigate the system from quite different perspectives. 
This new tool will enable the study of entropic-driven phenomena, like entropic forces, already found to be of great importance in many biological systems as flocks of birds~\cite{GiardinaEntropicEffects}. 

Furthermore, the general nature of the method encourages its application in many other systems of interest.

\section*{Acknowledgments}
The authors would like to thank Francesca Tria, Vito D.P. Servedio and Vittorio Loreto for useful insights, Andrea De Martino and Francesco Mauri for helpful discussions.

\appendix

\section{Mean-field MaxEnt}\label{app:MFME}
Here we derive \eqname~\eqref{eq:mf_sol}. In the mean-field approximation, only one body observables $\{x_i\}_{i = 1}^q$ are constrained, e.g. $\braket {x_i}$
is the average number of nodes in the $i$-th state. Now, \eqname~\eqref{h:sol} describes a non-interacting effective Hamiltonian. 
\beq
\braket{x_i} = \frac{1}{Z}\sum_{\sigma_1\cdots \sigma_M} \left(\sum_{k = 1}^M \delta_{\sigma_k, i}\right)\exp\left(- \sum_{h = 1}^q \lambda_h\sum_{k = 1}^M \delta_{\sigma_k, h}\right),
\eeq
where $Z$ is the normalization of the probability distribution and we identified the configuration $\vec \sigma$ with its site values:
\beq
\vec\sigma = \bpm \sigma_1 \\ \sigma_2 \\ \vdots\\\sigma_M\epm.
\eeq
The $\left(\sum_{k = 1}^M \delta_{\sigma_k, i}\right)$ is the application of the $x_i$ observable
on the $\vec \sigma$ state.
Since $x_i$ does not depend on the particular site $k$ we have:
\beq
\braket{x_i} = M \frac{e^{-\lambda_i}}{\sum_{h = 1}^q e^{-\lambda_h}}.
\label{eq:system}
\eeq
\eqname~\eqref{eq:system} defines a complete set of linear equations. They are dependent since we have the constraint:
\beq
\sum_{i = 1}^q \braket{x_i} = M
\eeq
It is straightforward to show that the most general solution of the system is given by:
\beq
\lambda_i = - \log\left(\frac{\braket{x_i}}{M}\right) + C
\label{eq:lambda}
\eeq
where $C$ is an arbitrary constant that does not affect any physical quantity. For sake of simplicity, in \eqname~\eqref{eq:mf_sol} we set $C = 0$.

\section{Importance sampling}\label{app:IS}
The minimization of \eqname~\eqref{chi2:new} is computationally expensive. In each step, the expected values of the observables for the trial set of $\lambda_i$ parameters must
be computed through a Monte Carlo-Metropolis integration. A natural extension of the Metropolis algorithm consists of re-weighting the extracted configurations  at each step.
This method takes the name of Importance Sampling (IS) and it has been widely applied in many physical applications~\cite{Torrie1977,Errea2014}.
It was  introduced in MaxEnt  by Broderick et al.~\cite{BialekIS}.

The average of an observable with an Hamiltonian $H'$ can be computed using a set uniform distributed configurations $\left\{c\right\}_{i = 1}^{N_c}$ as follows:
\beq
\braket A_{H'}=  \sum_{i = 1}^{N_c} \left(A(c_i)\frac{P_{H'}(c_i)}{P_H(c_i)}\right) P_H(c_i) = \braket{A \frac{P_{H'}}{P_H}}_H.
\eeq
This average can be computed using Monte Carlo integration on a set of Metropolis extracted  configurations $\left\{c'\right\}_{i = 1}^{N_c}$ with the $H$ Hamiltonian:
\begin{subequations}
  \begin{eqnarray}
    \braket A_{H'} &\approx& \frac{1}{Z(N_c)}\sum_{i = 1}^{N_c} A(c'_i)e^{-\beta \left[ H'(c'_i) - H(c'_i)\right]},\\
    Z(N_c) &=& \sum_{i = 1}^{N_c} e^{-\beta \left[ H'(c'_i) - H(c'_i)\right]}.
  \end{eqnarray}
\end{subequations}
Handling with large lattices, energy differences can be considerable, and the exponential term may give rise to numerical instabilities.
To correct these instabilities a constant factor $a$ can be added to both  exponential terms, equal to the maximum energy difference of all extracted configurations.

\begin{figure}
\includegraphics[width=\columnwidth]{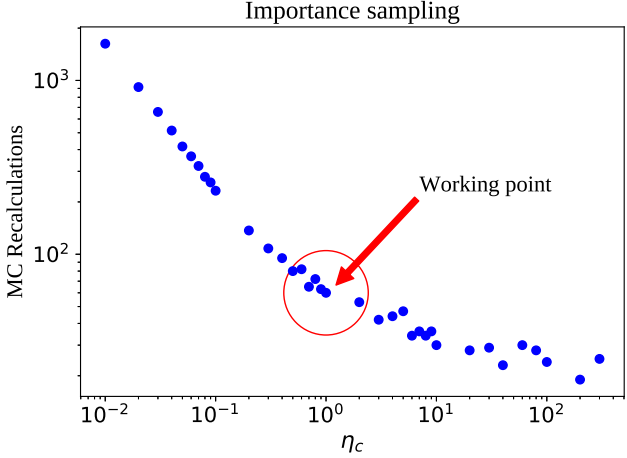}
\caption{(Color online) Number of Monte Carlo simulations to reach convergence as a function of the $\eta_c$ parameter~\eqref{eq:eta:p}.
The optimal working point lies in the marked region.
\label{fig:etaStudy}}
\end{figure}

Estimating the goodness of IS in MaxEnt implementation can be difficult. In fact, we lack the \emph{apriori} knowledge of the partition function of the
original probability distribution.
In order to upstage, the problem we implemented a new statistical evaluator for the MaxEnt algorithm and in general for IS Metropolis implementation.
At each step the total extracted configurations are divided into two random groups and the normalization factors are compared:
\beq
\label{eq:eta:p}
\eta =\left| \frac{Z'(N_c/2)}{Z''(N_c/2)} - 1\right|.
\eeq
If $\eta$ exceeds a critical value $\eta_c$, new configurations are extracted from the Metropolis algorithm.
\figurename~\ref{fig:etaStudy} shows the performance of IS vs the $\eta_c$ parameter. For $\eta_c\ll1$, \eqname~\eqref{eq:eta:p} is symmetric
with respect to $Z'$ and $Z''$. For higher values of $\eta_c$ the symmetry  is recovered by random shuffling the configurations at each step.

\section{Proof of the least maximum entropy principle}\label{app:least}
By definition, the MaxEnt entropy is always greater than the true entropy:
\beq
\label{eq:least}
S_{real} \le S_{ME}[\left\{\hat x_i\right\}]
\eeq
where $S_{real}$ is the real entropy of the system, while $S_{ME}[\left\{x_i\right\}]$ is the maximum entropy of all possible probability distributions that fix the set of $\hat x_i$ observables.

To  prove the least maximum entropy principle, a set of observables $\hat x_i$  must exist so that \eqname~\eqref{eq:least} is an equality.
In any finite size system, where configurations can be represented with a finite dimension vector, this is trivial, since it is possible to choose a set of observables $\hat y_i$ defined as:
\beq
\hat y_i(\vec \sigma) = \delta(\vec \sigma, \vec \sigma_i),
\eeq
where $\vec \sigma$ is the configuration on which the observables act, $\vec \sigma_i$ is a particular configuration associated with the observable $\hat y_i$ and $\delta$ is the Kronecker symbol.

If two probability distributions are different, then a configuration $\vec \sigma_i$ must exist so that their probabilities differ:
\beq
p_1(\vec \sigma_i) \neq p_2(\vec \sigma_i).
\eeq
The two distributions give two distinct expected values for the corresponding $\hat y_i$ observable:
\beq
\braket{y_i}_{p_1} = \sum_{j} p_1(\vec \sigma_j) \delta(\vec \sigma_j, \vec \sigma_i) = p_1(\vec \sigma_i) \neq p_2(\vec \sigma_i) = \braket{y_i}_{p_2}.
\eeq
The complete set of $\hat y_i$ fixes the probability distribution, so that
\beq
S_{real} = S_{ME}[\left\{\hat y_i\right\}].
\eeq
Indeed, single-site density and couple density can be written as linear combination of the complete set of $\hat y_i$ observables. Therefore, the introduction of more independent constraints assures the convergence of the MaxEnt entropy toward the real entropy.

\section{Dynamical Monte Carlo simulations of the EcoLat model}\label{app:dyn:mc}
Dynamical Monte Carlo allows us to simulate the EcoLat master equation (\eqname~\ref{eq:trans:mat}). Since the number of possible states is huge ($3^{M}$) it is impossible to numerically evolve the probability distribution. However, a stochastic solution of \eqname~\eqref{eq:trans:mat} is still possible: $N$ replicas of the system, extracted according to an initial distribution $P(\vec\sigma, 0)$, can be evolved according to the transition matrix $\Pi$. The obtained time-dependent ensemble
can be used to compute the averages of any observable as a function of time.

\begin{figure}
\centering
\includegraphics[width=\columnwidth]{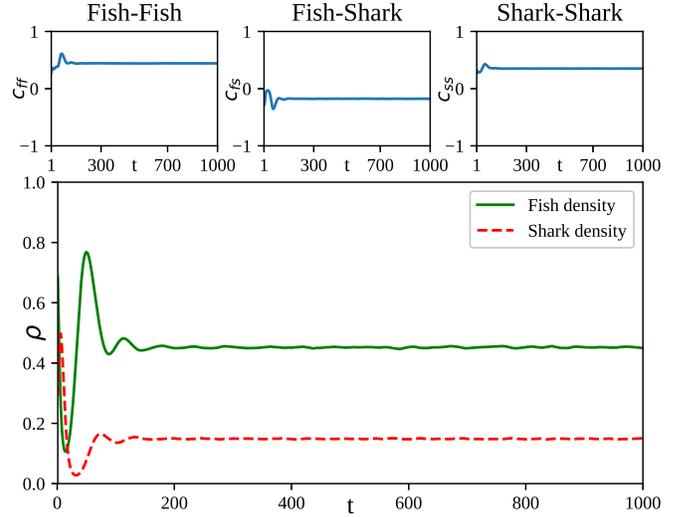}
\caption{\small Time evolution of the mean fish and shark densities and near-neighbour Pearson correlation coefficients computed on an ensemble 
composed by $100$ replicas of the system. The simulation
is prepared in an initial uniform distribution fixing the densities of preys and predators to $4/5$ and $1/5$ respectively.\label{fig:dynmont}}
\end{figure}

The EcoLat model  reaches the asymptotic steady-state distribution described by \eqname~\eqref{eq:equilibrium}. \figurename~\ref{fig:dynmont} shows the time dependence of the mean density of preys and predators as well as the near-neighbor correlation coefficients (\eqname~\ref{eq:corr:coeff}).

Note that, although the time evolution of the single system continues to oscillate in the steady-state, the correlation time is finite. This assures that, after a transient time, all the $N$ simulations are independent and distributed according to the $P(\vec \sigma)$ (\eqname~\ref{eq:equilibrium}).

\section{Ergodic hypothesis}\label{app:ergodic}
Here we discuss the stability of the steady-state distribution that depends on the ergodicity of the system~\cite{Huang}. 
Generally speaking, a system is ergodic if it can move  between any couple of points in the phase-space in a finite number of steps~\cite{Palmer1982}. EcoLat does not
satisfy this requirement, in fact, two traps are present in the phase-space: if the system gets a configuration without preys or predators, it always evolves toward an absorbing state. 
However, in simulations, once the system has reached the non-absorbing steady-state (\figurename~\ref{fig:dynmont}), it remains there during all the simulation
time. 
So, we can postulate a weak ergodic hypothesis, where we imagine to restrict the feasible phase-space excluding the absorbing traps. This is equivalent to neglecting
the elements of the $\Pi$ transition matrix that lead the system into the absorbing traps. Such an approximation makes sense if the lifetime $\tau$ of the meta-stable equilibrium is much larger than the typical time scales of the system. All the Dynamical Monte Carlo simulations computed in \figurename~\ref{fig:entropy} have $\tau \gg t_{max}$, where
$t_{max}$ is the maximum simulation time ($10^6$). 

\section{Simulation details}\label{app:par}
Phenotypic parameters for the EcoLat simulation of \figurename~\ref{fig::SF}  are:
$p_f^f = 0.2$, $p_f^m = 0.8$, $p_s^m = 0.7$, $p_s^f = 1$, $p_s^d = 0.3$. 
Phenotypic parameters for the EcoLat simulation of \figurename~\ref{fig:PTD} are the same apart for $p_s^m$ that assumes the values of 0.47, 0.70 and 0.90.

All simulations and computations  were carried out using authors handmade C and Python scripts. Python packages NumPy~\cite{van_der_Walt_2011}, SciPy~\cite{SciPy} and MatPlotlib~\cite{Hunter_2007} were utilized during analysis and figure realization.
 
\newpage
\bibliography{biblio}

\begin{thebibliography}{67}
\expandafter\ifx\csname natexlab\endcsname\relax\def\natexlab#1{#1}\fi
\expandafter\ifx\csname bibnamefont\endcsname\relax
  \def\bibnamefont#1{#1}\fi
\expandafter\ifx\csname bibfnamefont\endcsname\relax
  \def\bibfnamefont#1{#1}\fi
\expandafter\ifx\csname citenamefont\endcsname\relax
  \def\citenamefont#1{#1}\fi
\expandafter\ifx\csname url\endcsname\relax
  \def\url#1{\texttt{#1}}\fi
\expandafter\ifx\csname urlprefix\endcsname\relax\def\urlprefix{URL }\fi
\providecommand{\bibinfo}[2]{#2}
\providecommand{\eprint}[2][]{\url{#2}}

\bibitem[{\citenamefont{Johannsen}(1911)}]{Johannsen_1911}
\bibinfo{author}{\bibfnamefont{W.}~\bibnamefont{Johannsen}},
  \bibinfo{journal}{The American Naturalist} \textbf{\bibinfo{volume}{45}},
  \bibinfo{pages}{129} (\bibinfo{year}{1911}).

\bibitem[{\citenamefont{Churchill}(1974)}]{Churchill_1974}
\bibinfo{author}{\bibfnamefont{F.~B.} \bibnamefont{Churchill}},
  \bibinfo{journal}{Journal of the History of Biology}
  \textbf{\bibinfo{volume}{7}}, \bibinfo{pages}{5} (\bibinfo{year}{1974}).

\bibitem[{\citenamefont{Bialek}(2012)}]{bialek2012biophysics}
\bibinfo{author}{\bibfnamefont{W.}~\bibnamefont{Bialek}},
  \emph{\bibinfo{title}{Biophysics: Searching for Principles}}
  (\bibinfo{publisher}{Princeton University Press}, \bibinfo{year}{2012}).

\bibitem[{\citenamefont{Strong et~al.}(1998)\citenamefont{Strong, Koberle,
  de~Ruyter~van Steveninck, and Bialek}}]{Strong1998}
\bibinfo{author}{\bibfnamefont{S.~P.} \bibnamefont{Strong}},
  \bibinfo{author}{\bibfnamefont{R.}~\bibnamefont{Koberle}},
  \bibinfo{author}{\bibfnamefont{R.~R.} \bibnamefont{de~Ruyter~van
  Steveninck}}, \bibnamefont{and}
  \bibinfo{author}{\bibfnamefont{W.}~\bibnamefont{Bialek}},
  \bibinfo{journal}{Physical Review Letters} \textbf{\bibinfo{volume}{80}},
  \bibinfo{pages}{197} (\bibinfo{year}{1998}).

\bibitem[{\citenamefont{Nemenman et~al.}(2004)\citenamefont{Nemenman, Bialek,
  and {de Ruyter van Steveninck}}}]{Nemenman2004}
\bibinfo{author}{\bibfnamefont{I.}~\bibnamefont{Nemenman}},
  \bibinfo{author}{\bibfnamefont{W.}~\bibnamefont{Bialek}}, \bibnamefont{and}
  \bibinfo{author}{\bibfnamefont{R.}~\bibnamefont{{de Ruyter van Steveninck}}},
  \bibinfo{journal}{Physical Review E} \textbf{\bibinfo{volume}{69}}
  (\bibinfo{year}{2004}).

\bibitem[{\citenamefont{Bialek et~al.}(2012)\citenamefont{Bialek, Cavagna,
  Giardina, Mora, Silvestri, Viale, and Walczak}}]{Bialek2012}
\bibinfo{author}{\bibfnamefont{W.}~\bibnamefont{Bialek}},
  \bibinfo{author}{\bibfnamefont{A.}~\bibnamefont{Cavagna}},
  \bibinfo{author}{\bibfnamefont{I.}~\bibnamefont{Giardina}},
  \bibinfo{author}{\bibfnamefont{T.}~\bibnamefont{Mora}},
  \bibinfo{author}{\bibfnamefont{E.}~\bibnamefont{Silvestri}},
  \bibinfo{author}{\bibfnamefont{M.}~\bibnamefont{Viale}}, \bibnamefont{and}
  \bibinfo{author}{\bibfnamefont{A.~M.} \bibnamefont{Walczak}},
  \bibinfo{journal}{Proceedings of the National Academy of Sciences}
  \textbf{\bibinfo{volume}{109}}, \bibinfo{pages}{4786} (\bibinfo{year}{2012}).

\bibitem[{\citenamefont{Castellana et~al.}(2016)\citenamefont{Castellana,
  Bialek, Cavagna, and Giardina}}]{GiardinaEntropicEffects}
\bibinfo{author}{\bibfnamefont{M.}~\bibnamefont{Castellana}},
  \bibinfo{author}{\bibfnamefont{W.}~\bibnamefont{Bialek}},
  \bibinfo{author}{\bibfnamefont{A.}~\bibnamefont{Cavagna}}, \bibnamefont{and}
  \bibinfo{author}{\bibfnamefont{I.}~\bibnamefont{Giardina}},
  \bibinfo{journal}{Phys. Rev. E} \textbf{\bibinfo{volume}{93}}
  (\bibinfo{year}{2016}).

\bibitem[{\citenamefont{Bialek et~al.}(2001)\citenamefont{Bialek, Nemenman, and
  Tishby}}]{Bialek2001}
\bibinfo{author}{\bibfnamefont{W.}~\bibnamefont{Bialek}},
  \bibinfo{author}{\bibfnamefont{I.}~\bibnamefont{Nemenman}}, \bibnamefont{and}
  \bibinfo{author}{\bibfnamefont{N.}~\bibnamefont{Tishby}},
  \bibinfo{journal}{Neural Computation} \textbf{\bibinfo{volume}{13}},
  \bibinfo{pages}{2409} (\bibinfo{year}{2001}).

\bibitem[{\citenamefont{Shannon}(1951{\natexlab{a}})}]{Shannon1951}
\bibinfo{author}{\bibfnamefont{C.~E.} \bibnamefont{Shannon}},
  \bibinfo{journal}{Bell System Technical Journal}
  \textbf{\bibinfo{volume}{30}}, \bibinfo{pages}{50}
  (\bibinfo{year}{1951}{\natexlab{a}}).

\bibitem[{\citenamefont{Kelly}(1956)}]{Kelly1956}
\bibinfo{author}{\bibfnamefont{J.~L.} \bibnamefont{Kelly}},
  \bibinfo{journal}{Bell System Technical Journal}
  \textbf{\bibinfo{volume}{35}}, \bibinfo{pages}{917} (\bibinfo{year}{1956}).

\bibitem[{\citenamefont{Philippatos and Wilson}(1972)}]{Philippatos1972}
\bibinfo{author}{\bibfnamefont{G.~C.} \bibnamefont{Philippatos}}
  \bibnamefont{and} \bibinfo{author}{\bibfnamefont{C.~J.}
  \bibnamefont{Wilson}}, \bibinfo{journal}{Applied Economics}
  \textbf{\bibinfo{volume}{4}}, \bibinfo{pages}{209} (\bibinfo{year}{1972}).

\bibitem[{\citenamefont{{De Martino} et~al.}(2017)\citenamefont{{De Martino},
  Capuani, and {De Martino}}}]{DeMartino2017}
\bibinfo{author}{\bibfnamefont{D.}~\bibnamefont{{De Martino}}},
  \bibinfo{author}{\bibfnamefont{F.}~\bibnamefont{Capuani}}, \bibnamefont{and}
  \bibinfo{author}{\bibfnamefont{A.}~\bibnamefont{{De Martino}}},
  \bibinfo{journal}{Physical Review E} \textbf{\bibinfo{volume}{96}}
  (\bibinfo{year}{2017}).

\bibitem[{\citenamefont{{De Martino} et~al.}(2016)\citenamefont{{De Martino},
  Capuani, and {De Martino}}}]{2DeMartino2017}
\bibinfo{author}{\bibfnamefont{D.}~\bibnamefont{{De Martino}}},
  \bibinfo{author}{\bibfnamefont{F.}~\bibnamefont{Capuani}}, \bibnamefont{and}
  \bibinfo{author}{\bibfnamefont{A.}~\bibnamefont{{De Martino}}},
  \bibinfo{journal}{Physical Biology} \textbf{\bibinfo{volume}{13}}
  (\bibinfo{year}{2016}).

\bibitem[{\citenamefont{Kussell}(2005)}]{Kussell2005}
\bibinfo{author}{\bibfnamefont{E.}~\bibnamefont{Kussell}},
  \bibinfo{journal}{Science} \textbf{\bibinfo{volume}{309}},
  \bibinfo{pages}{2075} (\bibinfo{year}{2005}).

\bibitem[{\citenamefont{Weigt et~al.}(2008)\citenamefont{Weigt, White,
  Szurmant, Hoch, and Hwa}}]{Weigt2008}
\bibinfo{author}{\bibfnamefont{M.}~\bibnamefont{Weigt}},
  \bibinfo{author}{\bibfnamefont{R.~A.} \bibnamefont{White}},
  \bibinfo{author}{\bibfnamefont{H.}~\bibnamefont{Szurmant}},
  \bibinfo{author}{\bibfnamefont{J.~A.} \bibnamefont{Hoch}}, \bibnamefont{and}
  \bibinfo{author}{\bibfnamefont{T.}~\bibnamefont{Hwa}},
  \bibinfo{journal}{Proceedings of the National Academy of Sciences}
  \textbf{\bibinfo{volume}{106}}, \bibinfo{pages}{67} (\bibinfo{year}{2008}).

\bibitem[{\citenamefont{Tansley}(1935)}]{Tansley1935}
\bibinfo{author}{\bibfnamefont{A.~G.} \bibnamefont{Tansley}},
  \bibinfo{journal}{Ecology} \textbf{\bibinfo{volume}{16}},
  \bibinfo{pages}{284} (\bibinfo{year}{1935}).

\bibitem[{\citenamefont{Allee}(1934)}]{Allee1934}
\bibinfo{author}{\bibfnamefont{W.~C.} \bibnamefont{Allee}},
  \bibinfo{journal}{Ecological Monographs} \textbf{\bibinfo{volume}{4}},
  \bibinfo{pages}{541} (\bibinfo{year}{1934}).

\bibitem[{\citenamefont{Chapin et~al.}(2011)\citenamefont{Chapin, Matson, and
  Vitousek}}]{Chapin2011}
\bibinfo{author}{\bibfnamefont{F.~S.} \bibnamefont{Chapin}},
  \bibinfo{author}{\bibfnamefont{P.~A.} \bibnamefont{Matson}},
  \bibnamefont{and} \bibinfo{author}{\bibfnamefont{P.~M.}
  \bibnamefont{Vitousek}}, \emph{\bibinfo{title}{Principles of Terrestrial
  Ecosystem Ecology}} (\bibinfo{publisher}{Springer New York},
  \bibinfo{year}{2011}).

\bibitem[{\citenamefont{Cooper and Blumstein}(2015)}]{cooper2015escaping}
\bibinfo{author}{\bibfnamefont{W.}~\bibnamefont{Cooper}} \bibnamefont{and}
  \bibinfo{author}{\bibfnamefont{D.}~\bibnamefont{Blumstein}},
  \emph{\bibinfo{title}{Escaping From Predators: An Integrative View of Escape
  Decisions}} (\bibinfo{publisher}{Cambridge University Press},
  \bibinfo{year}{2015}).

\bibitem[{\citenamefont{Belgrad and Griffen}(2016)}]{Belgrad_2016}
\bibinfo{author}{\bibfnamefont{B.~A.} \bibnamefont{Belgrad}} \bibnamefont{and}
  \bibinfo{author}{\bibfnamefont{B.~D.} \bibnamefont{Griffen}},
  \bibinfo{journal}{Proceedings of the Royal Society B: Biological Sciences}
  \textbf{\bibinfo{volume}{283}} (\bibinfo{year}{2016}).

\bibitem[{\citenamefont{Olson et~al.}(2013)\citenamefont{Olson, Hintze, Dyer,
  Knoester, and Adami}}]{Olson_2013}
\bibinfo{author}{\bibfnamefont{R.~S.} \bibnamefont{Olson}},
  \bibinfo{author}{\bibfnamefont{A.}~\bibnamefont{Hintze}},
  \bibinfo{author}{\bibfnamefont{F.~C.} \bibnamefont{Dyer}},
  \bibinfo{author}{\bibfnamefont{D.~B.} \bibnamefont{Knoester}},
  \bibnamefont{and} \bibinfo{author}{\bibfnamefont{C.}~\bibnamefont{Adami}},
  \bibinfo{journal}{Journal of The Royal Society Interface}
  \textbf{\bibinfo{volume}{10}} (\bibinfo{year}{2013}).

\bibitem[{\citenamefont{Swetnam and Betancourt}(1998)}]{disturbance}
\bibinfo{author}{\bibfnamefont{T.~W.} \bibnamefont{Swetnam}} \bibnamefont{and}
  \bibinfo{author}{\bibfnamefont{J.~L.} \bibnamefont{Betancourt}},
  \bibinfo{journal}{Journal of Climate} \textbf{\bibinfo{volume}{11}},
  \bibinfo{pages}{3128} (\bibinfo{year}{1998}).

\bibitem[{\citenamefont{Kindlmann et~al.}(2015)\citenamefont{Kindlmann, Yasuda,
  Kajita, Sato, and Dixon}}]{Kindlmann2015}
\bibinfo{author}{\bibfnamefont{P.}~\bibnamefont{Kindlmann}},
  \bibinfo{author}{\bibfnamefont{H.}~\bibnamefont{Yasuda}},
  \bibinfo{author}{\bibfnamefont{Y.}~\bibnamefont{Kajita}},
  \bibinfo{author}{\bibfnamefont{S.}~\bibnamefont{Sato}}, \bibnamefont{and}
  \bibinfo{author}{\bibfnamefont{A.~F.~G.} \bibnamefont{Dixon}},
  \bibinfo{journal}{Frontiers in Ecology and Evolution}
  \textbf{\bibinfo{volume}{3}} (\bibinfo{year}{2015}).

\bibitem[{\citenamefont{Bax}(1998)}]{Bax1998}
\bibinfo{author}{\bibfnamefont{N.}~\bibnamefont{Bax}}, \bibinfo{journal}{{ICES}
  Journal of Marine Science} \textbf{\bibinfo{volume}{55}},
  \bibinfo{pages}{997} (\bibinfo{year}{1998}).

\bibitem[{\citenamefont{Lotka}(1920)}]{Lotka}
\bibinfo{author}{\bibfnamefont{A.}~\bibnamefont{Lotka}},
  \bibinfo{journal}{Proc. Natl. Acad. Sci. U.S}  (\bibinfo{year}{1920}).

\bibitem[{\citenamefont{Volterra}(1931)}]{volterra}
\bibinfo{author}{\bibfnamefont{V.}~\bibnamefont{Volterra}},
  \bibinfo{journal}{McGraw-Hill}  (\bibinfo{year}{1931}).

\bibitem[{\citenamefont{Holmes et~al.}(1994)\citenamefont{Holmes, Lewis, Banks,
  and Veit}}]{Holmes_1994}
\bibinfo{author}{\bibfnamefont{E.~E.} \bibnamefont{Holmes}},
  \bibinfo{author}{\bibfnamefont{M.~A.} \bibnamefont{Lewis}},
  \bibinfo{author}{\bibfnamefont{J.~E.} \bibnamefont{Banks}}, \bibnamefont{and}
  \bibinfo{author}{\bibfnamefont{R.~R.} \bibnamefont{Veit}},
  \bibinfo{journal}{Ecology} \textbf{\bibinfo{volume}{75}}, \bibinfo{pages}{17}
  (\bibinfo{year}{1994}).

\bibitem[{\citenamefont{Kuto and Yamada}(2004)}]{Kuto_2004}
\bibinfo{author}{\bibfnamefont{K.}~\bibnamefont{Kuto}} \bibnamefont{and}
  \bibinfo{author}{\bibfnamefont{Y.}~\bibnamefont{Yamada}},
  \bibinfo{journal}{Journal of Differential Equations}
  \textbf{\bibinfo{volume}{197}}, \bibinfo{pages}{315} (\bibinfo{year}{2004}).

\bibitem[{\citenamefont{Ferreira et~al.}(2013)\citenamefont{Ferreira, Salazar,
  and Tabares}}]{Ferreira_2013}
\bibinfo{author}{\bibfnamefont{J.~D.} \bibnamefont{Ferreira}},
  \bibinfo{author}{\bibfnamefont{C.~A.~T.} \bibnamefont{Salazar}},
  \bibnamefont{and} \bibinfo{author}{\bibfnamefont{P.~C.}
  \bibnamefont{Tabares}}, \bibinfo{journal}{Nonlinear Analysis: Real World
  Applications} \textbf{\bibinfo{volume}{14}}, \bibinfo{pages}{536}
  (\bibinfo{year}{2013}).

\bibitem[{\citenamefont{Satulovsky and Tom{\'{e}}}(1994)}]{Satulovsky_1994}
\bibinfo{author}{\bibfnamefont{J.~E.} \bibnamefont{Satulovsky}}
  \bibnamefont{and}
  \bibinfo{author}{\bibfnamefont{T.}~\bibnamefont{Tom{\'{e}}}},
  \bibinfo{journal}{Physical Review E} \textbf{\bibinfo{volume}{49}},
  \bibinfo{pages}{5073} (\bibinfo{year}{1994}).

\bibitem[{\citenamefont{Lipowski}(1999)}]{Lipowski_1999}
\bibinfo{author}{\bibfnamefont{A.}~\bibnamefont{Lipowski}},
  \bibinfo{journal}{Physical Review E} \textbf{\bibinfo{volume}{60}},
  \bibinfo{pages}{5179} (\bibinfo{year}{1999}).

\bibitem[{\citenamefont{Dobramysl et~al.}(2018)\citenamefont{Dobramysl,
  Mobilia, Pleimling, and T\"{a}uber}}]{Dobramysl2018}
\bibinfo{author}{\bibfnamefont{U.}~\bibnamefont{Dobramysl}},
  \bibinfo{author}{\bibfnamefont{M.}~\bibnamefont{Mobilia}},
  \bibinfo{author}{\bibfnamefont{M.}~\bibnamefont{Pleimling}},
  \bibnamefont{and} \bibinfo{author}{\bibfnamefont{U.~C.}
  \bibnamefont{T\"{a}uber}}, \bibinfo{journal}{Journal of Physics A:
  Mathematical and Theoretical} \textbf{\bibinfo{volume}{51}}
  (\bibinfo{year}{2018}).

\bibitem[{\citenamefont{Dewdney}(1984)}]{dewdney}
\bibinfo{author}{\bibfnamefont{A.~K.} \bibnamefont{Dewdney}},
  \bibinfo{journal}{Scientific American} \textbf{\bibinfo{volume}{251}},
  \bibinfo{pages}{14} (\bibinfo{year}{1984}).

\bibitem[{\citenamefont{Mobilia et~al.}(2006)\citenamefont{Mobilia, Georgiev,
  and T\"{a}uber}}]{Mobilia2006}
\bibinfo{author}{\bibfnamefont{M.}~\bibnamefont{Mobilia}},
  \bibinfo{author}{\bibfnamefont{I.~T.} \bibnamefont{Georgiev}},
  \bibnamefont{and} \bibinfo{author}{\bibfnamefont{U.~C.}
  \bibnamefont{T\"{a}uber}}, \bibinfo{journal}{Physical Review E}
  \textbf{\bibinfo{volume}{73}} (\bibinfo{year}{2006}).

\bibitem[{\citenamefont{Antal et~al.}(2001)\citenamefont{Antal, Droz, Lipowski,
  and {\'{O}}dor}}]{Antal2001}
\bibinfo{author}{\bibfnamefont{T.}~\bibnamefont{Antal}},
  \bibinfo{author}{\bibfnamefont{M.}~\bibnamefont{Droz}},
  \bibinfo{author}{\bibfnamefont{A.}~\bibnamefont{Lipowski}}, \bibnamefont{and}
  \bibinfo{author}{\bibfnamefont{G.}~\bibnamefont{{\'{O}}dor}},
  \bibinfo{journal}{Physical Review E} \textbf{\bibinfo{volume}{64}}
  (\bibinfo{year}{2001}).

\bibitem[{\citenamefont{T\"{a}uber}(2011)}]{Tuber2011}
\bibinfo{author}{\bibfnamefont{U.~C.} \bibnamefont{T\"{a}uber}},
  \bibinfo{journal}{Journal of Physics: Conference Series}
  \textbf{\bibinfo{volume}{319}}, \bibinfo{pages}{012019}
  (\bibinfo{year}{2011}).

\bibitem[{\citenamefont{Nemenman et~al.}(2002)\citenamefont{Nemenman, Shafee,
  and Bialek}}]{nemenman2002entropy}
\bibinfo{author}{\bibfnamefont{I.}~\bibnamefont{Nemenman}},
  \bibinfo{author}{\bibfnamefont{F.}~\bibnamefont{Shafee}}, \bibnamefont{and}
  \bibinfo{author}{\bibfnamefont{W.}~\bibnamefont{Bialek}}, in
  \emph{\bibinfo{booktitle}{Advances in neural information processing systems}}
  (\bibinfo{year}{2002}), pp. \bibinfo{pages}{471--478}.

\bibitem[{\citenamefont{Shannon}(1951{\natexlab{b}})}]{shannon_game}
\bibinfo{author}{\bibfnamefont{C.~E.} \bibnamefont{Shannon}},
  \bibinfo{journal}{Bell system technical journal}
  (\bibinfo{year}{1951}{\natexlab{b}}).

\bibitem[{\citenamefont{Jaynes}(1957)}]{Jaynes1957}
\bibinfo{author}{\bibfnamefont{E.~T.} \bibnamefont{Jaynes}},
  \bibinfo{journal}{Physical Review} \textbf{\bibinfo{volume}{106}},
  \bibinfo{pages}{620} (\bibinfo{year}{1957}).

\bibitem[{\citenamefont{Cavagna et~al.}(2014)\citenamefont{Cavagna, Giardina,
  Ginelli, Mora, Piovani, Tavarone, and Walczak}}]{Cavagna2014}
\bibinfo{author}{\bibfnamefont{A.}~\bibnamefont{Cavagna}},
  \bibinfo{author}{\bibfnamefont{I.}~\bibnamefont{Giardina}},
  \bibinfo{author}{\bibfnamefont{F.}~\bibnamefont{Ginelli}},
  \bibinfo{author}{\bibfnamefont{T.}~\bibnamefont{Mora}},
  \bibinfo{author}{\bibfnamefont{D.}~\bibnamefont{Piovani}},
  \bibinfo{author}{\bibfnamefont{R.}~\bibnamefont{Tavarone}}, \bibnamefont{and}
  \bibinfo{author}{\bibfnamefont{A.~M.} \bibnamefont{Walczak}},
  \bibinfo{journal}{Physical Review E} \textbf{\bibinfo{volume}{89}}
  (\bibinfo{year}{2014}).

\bibitem[{\citenamefont{Mora et~al.}(2010)\citenamefont{Mora, Walczak, Bialek,
  and Callan}}]{Mora2010}
\bibinfo{author}{\bibfnamefont{T.}~\bibnamefont{Mora}},
  \bibinfo{author}{\bibfnamefont{A.~M.} \bibnamefont{Walczak}},
  \bibinfo{author}{\bibfnamefont{W.}~\bibnamefont{Bialek}}, \bibnamefont{and}
  \bibinfo{author}{\bibfnamefont{C.~G.} \bibnamefont{Callan}},
  \bibinfo{journal}{Proceedings of the National Academy of Sciences}
  \textbf{\bibinfo{volume}{107}}, \bibinfo{pages}{5405} (\bibinfo{year}{2010}).

\bibitem[{\citenamefont{Schneidman et~al.}(2006)\citenamefont{Schneidman,
  Berry, Segev, and Bialek}}]{Schneidman2006}
\bibinfo{author}{\bibfnamefont{E.}~\bibnamefont{Schneidman}},
  \bibinfo{author}{\bibfnamefont{M.~J.} \bibnamefont{Berry}},
  \bibinfo{author}{\bibfnamefont{R.}~\bibnamefont{Segev}}, \bibnamefont{and}
  \bibinfo{author}{\bibfnamefont{W.}~\bibnamefont{Bialek}},
  \bibinfo{journal}{Nature} \textbf{\bibinfo{volume}{440}},
  \bibinfo{pages}{1007} (\bibinfo{year}{2006}).

\bibitem[{\citenamefont{Shannon}(1948)}]{Shannon1948}
\bibinfo{author}{\bibfnamefont{C.~E.} \bibnamefont{Shannon}},
  \bibinfo{journal}{Bell System Technical Journal}
  \textbf{\bibinfo{volume}{27}}, \bibinfo{pages}{379} (\bibinfo{year}{1948}).

\bibitem[{\citenamefont{Fano}(1949)}]{Fano1949}
\bibinfo{author}{\bibfnamefont{R.~M.} \bibnamefont{Fano}},
  \emph{\bibinfo{title}{The trasmission of Information}}
  (\bibinfo{publisher}{Cambrige: Massachusetts Institute of Technology,
  Research Laboratory of Electronics}, \bibinfo{year}{1949}).

\bibitem[{\citenamefont{Lockless}(1999)}]{Lockless_1999}
\bibinfo{author}{\bibfnamefont{S.~W.} \bibnamefont{Lockless}},
  \bibinfo{journal}{Science} \textbf{\bibinfo{volume}{286}},
  \bibinfo{pages}{295} (\bibinfo{year}{1999}).

\bibitem[{\citenamefont{Socolich et~al.}(2005)\citenamefont{Socolich, Lockless,
  Russ, Lee, Gardner, and Ranganathan}}]{Socolich_2005}
\bibinfo{author}{\bibfnamefont{M.}~\bibnamefont{Socolich}},
  \bibinfo{author}{\bibfnamefont{S.~W.} \bibnamefont{Lockless}},
  \bibinfo{author}{\bibfnamefont{W.~P.} \bibnamefont{Russ}},
  \bibinfo{author}{\bibfnamefont{H.}~\bibnamefont{Lee}},
  \bibinfo{author}{\bibfnamefont{K.~H.} \bibnamefont{Gardner}},
  \bibnamefont{and}
  \bibinfo{author}{\bibfnamefont{R.}~\bibnamefont{Ranganathan}},
  \bibinfo{journal}{Nature} \textbf{\bibinfo{volume}{437}},
  \bibinfo{pages}{512} (\bibinfo{year}{2005}).

\bibitem[{\citenamefont{Russ et~al.}(2005)\citenamefont{Russ, Lowery, Mishra,
  Yaffe, and Ranganathan}}]{Russ_2005}
\bibinfo{author}{\bibfnamefont{W.~P.} \bibnamefont{Russ}},
  \bibinfo{author}{\bibfnamefont{D.~M.} \bibnamefont{Lowery}},
  \bibinfo{author}{\bibfnamefont{P.}~\bibnamefont{Mishra}},
  \bibinfo{author}{\bibfnamefont{M.~B.} \bibnamefont{Yaffe}}, \bibnamefont{and}
  \bibinfo{author}{\bibfnamefont{R.}~\bibnamefont{Ranganathan}},
  \bibinfo{journal}{Nature} \textbf{\bibinfo{volume}{437}},
  \bibinfo{pages}{579} (\bibinfo{year}{2005}).

\bibitem[{\citenamefont{{Bialek} and {Ranganathan}}(2007)}]{2007arXiv}
\bibinfo{author}{\bibfnamefont{W.}~\bibnamefont{{Bialek}}} \bibnamefont{and}
  \bibinfo{author}{\bibfnamefont{R.}~\bibnamefont{{Ranganathan}}},
  \bibinfo{journal}{ArXiv e-prints}  (\bibinfo{year}{2007}).

\bibitem[{\citenamefont{Pearson}(1901)}]{Pearson_1901}
\bibinfo{author}{\bibfnamefont{K.}~\bibnamefont{Pearson}},
  \bibinfo{journal}{Philosophical Magazine Series 6}
  \textbf{\bibinfo{volume}{2}}, \bibinfo{pages}{559} (\bibinfo{year}{1901}).

\bibitem[{\citenamefont{Hotelling}(1933)}]{Hotelling_1933}
\bibinfo{author}{\bibfnamefont{H.}~\bibnamefont{Hotelling}},
  \bibinfo{journal}{Journal of Educational Psychology}
  \textbf{\bibinfo{volume}{24}}, \bibinfo{pages}{417} (\bibinfo{year}{1933}).

\bibitem[{\citenamefont{Nash}(1985)}]{Nash1985}
\bibinfo{author}{\bibfnamefont{S.~G.} \bibnamefont{Nash}},
  \bibinfo{journal}{{SIAM} Journal on Scientific and Statistical Computing}
  \textbf{\bibinfo{volume}{6}}, \bibinfo{pages}{599} (\bibinfo{year}{1985}).

\bibitem[{\citenamefont{Potts and Domb}(1952)}]{Potts_1952}
\bibinfo{author}{\bibfnamefont{R.~B.} \bibnamefont{Potts}} \bibnamefont{and}
  \bibinfo{author}{\bibfnamefont{C.}~\bibnamefont{Domb}},
  \bibinfo{journal}{Mathematical Proceedings of the Cambridge Philosophical
  Society} \textbf{\bibinfo{volume}{48}}, \bibinfo{pages}{106}
  (\bibinfo{year}{1952}).

\bibitem[{\citenamefont{Zia and Schmittmann}(2007)}]{Zia2007}
\bibinfo{author}{\bibfnamefont{R.~K.~P.} \bibnamefont{Zia}} \bibnamefont{and}
  \bibinfo{author}{\bibfnamefont{B.}~\bibnamefont{Schmittmann}},
  \bibinfo{journal}{Journal of Statistical Mechanics: Theory and Experiment}
  \textbf{\bibinfo{volume}{2007}}, \bibinfo{pages}{P07012}
  (\bibinfo{year}{2007}).

\bibitem[{\citenamefont{Sutherland and Jacobs}(1994)}]{Sutherland1994}
\bibinfo{author}{\bibfnamefont{B.}~\bibnamefont{Sutherland}} \bibnamefont{and}
  \bibinfo{author}{\bibfnamefont{A.}~\bibnamefont{Jacobs}},
  \bibinfo{journal}{Complex System} \textbf{\bibinfo{volume}{8}},
  \bibinfo{pages}{385} (\bibinfo{year}{1994}).

\bibitem[{\citenamefont{Pascual et~al.}(2002)\citenamefont{Pascual, Roy,
  Guichard, and Flierl}}]{Pascual2002}
\bibinfo{author}{\bibfnamefont{M.}~\bibnamefont{Pascual}},
  \bibinfo{author}{\bibfnamefont{M.}~\bibnamefont{Roy}},
  \bibinfo{author}{\bibfnamefont{F.}~\bibnamefont{Guichard}}, \bibnamefont{and}
  \bibinfo{author}{\bibfnamefont{G.}~\bibnamefont{Flierl}},
  \bibinfo{journal}{Philosophical Transactions of the Royal Society B:
  Biological Sciences} \textbf{\bibinfo{volume}{357}}, \bibinfo{pages}{657}
  (\bibinfo{year}{2002}).

\bibitem[{\citenamefont{Mastromatteo and Marsili}(2011)}]{Mastromatteo2011}
\bibinfo{author}{\bibfnamefont{I.}~\bibnamefont{Mastromatteo}}
  \bibnamefont{and} \bibinfo{author}{\bibfnamefont{M.}~\bibnamefont{Marsili}},
  \bibinfo{journal}{Journal of Statistical Mechanics: Theory and Experiment}
  \textbf{\bibinfo{volume}{2011}}, \bibinfo{pages}{P10012}
  (\bibinfo{year}{2011}).

\bibitem[{\citenamefont{Parker and Kamenev}(2009)}]{PhysRevE.80.021129}
\bibinfo{author}{\bibfnamefont{M.}~\bibnamefont{Parker}} \bibnamefont{and}
  \bibinfo{author}{\bibfnamefont{A.}~\bibnamefont{Kamenev}},
  \bibinfo{journal}{Phys. Rev. E} \textbf{\bibinfo{volume}{80}},
  \bibinfo{pages}{021129} (\bibinfo{year}{2009}).

\bibitem[{\citenamefont{Dobrinevski and Frey}(2012)}]{PhysRevE.85.051903}
\bibinfo{author}{\bibfnamefont{A.}~\bibnamefont{Dobrinevski}} \bibnamefont{and}
  \bibinfo{author}{\bibfnamefont{E.}~\bibnamefont{Frey}},
  \bibinfo{journal}{Phys. Rev. E} \textbf{\bibinfo{volume}{85}},
  \bibinfo{pages}{051903} (\bibinfo{year}{2012}).

\bibitem[{\citenamefont{Mobilia et~al.}(2007)\citenamefont{Mobilia, Georgiev,
  and T{\"a}uber}}]{Mobilia2007p}
\bibinfo{author}{\bibfnamefont{M.}~\bibnamefont{Mobilia}},
  \bibinfo{author}{\bibfnamefont{I.~T.} \bibnamefont{Georgiev}},
  \bibnamefont{and} \bibinfo{author}{\bibfnamefont{U.~C.}
  \bibnamefont{T{\"a}uber}}, \bibinfo{journal}{Journal of Statistical Physics}
  \textbf{\bibinfo{volume}{128}}, \bibinfo{pages}{447} (\bibinfo{year}{2007}).

\bibitem[{\citenamefont{Torrie and Valleau}(1977)}]{Torrie1977}
\bibinfo{author}{\bibfnamefont{G.}~\bibnamefont{Torrie}} \bibnamefont{and}
  \bibinfo{author}{\bibfnamefont{J.}~\bibnamefont{Valleau}},
  \bibinfo{journal}{Journal of Computational Physics}
  \textbf{\bibinfo{volume}{23}}, \bibinfo{pages}{187} (\bibinfo{year}{1977}).

\bibitem[{\citenamefont{Errea et~al.}(2014)\citenamefont{Errea, Calandra, and
  Mauri}}]{Errea2014}
\bibinfo{author}{\bibfnamefont{I.}~\bibnamefont{Errea}},
  \bibinfo{author}{\bibfnamefont{M.}~\bibnamefont{Calandra}}, \bibnamefont{and}
  \bibinfo{author}{\bibfnamefont{F.}~\bibnamefont{Mauri}},
  \bibinfo{journal}{Physical Review B} \textbf{\bibinfo{volume}{89}}
  (\bibinfo{year}{2014}).

\bibitem[{\citenamefont{{Broderick} et~al.}(2007)\citenamefont{{Broderick},
  {Dudik}, {Tkacik}, {Schapire}, and {Bialek}}}]{BialekIS}
\bibinfo{author}{\bibfnamefont{T.}~\bibnamefont{{Broderick}}},
  \bibinfo{author}{\bibfnamefont{M.}~\bibnamefont{{Dudik}}},
  \bibinfo{author}{\bibfnamefont{G.}~\bibnamefont{{Tkacik}}},
  \bibinfo{author}{\bibfnamefont{R.~E.} \bibnamefont{{Schapire}}},
  \bibnamefont{and} \bibinfo{author}{\bibfnamefont{W.}~\bibnamefont{{Bialek}}},
  \bibinfo{journal}{ArXiv e-prints}  (\bibinfo{year}{2007}),
  \eprint{0712.2437}.

\bibitem[{\citenamefont{Huang}(208)}]{Huang}
\bibinfo{author}{\bibfnamefont{K.}~\bibnamefont{Huang}},
  \emph{\bibinfo{title}{Statistical Mechanics}} (\bibinfo{publisher}{Wiley},
  \bibinfo{year}{208}).

\bibitem[{\citenamefont{Palmer}(1982)}]{Palmer1982}
\bibinfo{author}{\bibfnamefont{R.}~\bibnamefont{Palmer}},
  \bibinfo{journal}{Advances in Physics} \textbf{\bibinfo{volume}{31}},
  \bibinfo{pages}{669} (\bibinfo{year}{1982}).

\bibitem[{\citenamefont{{van der Walt} et~al.}(2011)\citenamefont{{van der
  Walt}, Colbert, and Varoquaux}}]{van_der_Walt_2011}
\bibinfo{author}{\bibfnamefont{S.}~\bibnamefont{{van der Walt}}},
  \bibinfo{author}{\bibfnamefont{S.~C.} \bibnamefont{Colbert}},
  \bibnamefont{and}
  \bibinfo{author}{\bibfnamefont{G.}~\bibnamefont{Varoquaux}},
  \bibinfo{journal}{Computing in Science {\&} Engineering}
  \textbf{\bibinfo{volume}{13}}, \bibinfo{pages}{22} (\bibinfo{year}{2011}).

\bibitem[{\citenamefont{Jones et~al.}(2001)\citenamefont{Jones, Oliphant,
  Peterson et~al.}}]{SciPy}
\bibinfo{author}{\bibfnamefont{E.}~\bibnamefont{Jones}},
  \bibinfo{author}{\bibfnamefont{T.}~\bibnamefont{Oliphant}},
  \bibinfo{author}{\bibfnamefont{P.}~\bibnamefont{Peterson}},
  \bibnamefont{et~al.}, \emph{\bibinfo{title}{{SciPy}: Open source scientific
  tools for {Python}}} (\bibinfo{year}{2001}).

\bibitem[{\citenamefont{Hunter}(2007)}]{Hunter_2007}
\bibinfo{author}{\bibfnamefont{J.~D.} \bibnamefont{Hunter}},
  \bibinfo{journal}{Computing In Science \& Engineering}
  \textbf{\bibinfo{volume}{9}}, \bibinfo{pages}{90} (\bibinfo{year}{2007}).

\end{thebibliography}


\end{document}